\documentclass[10pt, preprint]{emulateapj}
\usepackage{graphicx}
\usepackage{natbib}
\usepackage{amsmath}
\usepackage{amssymb}
\usepackage{fancyhdr}
\usepackage{latexsym}
\usepackage{color}
\usepackage{txfonts}

\begin{document}

\title{Solar Flare Chromospheric Line Emission: Comparison Between IBIS High-resolution Observations and Radiative Hydrodynamic Simulations}

\author{Fatima Rubio da Costa\altaffilmark{1}, Lucia Kleint\altaffilmark{2}, Vah\'{e} Petrosian\altaffilmark{1}, Alberto Sainz Dalda\altaffilmark{3} \\ and Wei Liu\altaffilmark{4}$^,\,$\altaffilmark{5}}
\altaffiltext{1}{Department of Physics, Stanford University, Stanford, CA 94305, USA; Email: frubio@stanford.edu}
\altaffiltext{2}{University of Applied Sciences and Arts Northwestern Switzerland, 5210 Windisch, Switzerland}
\altaffiltext{3}{Stanford-Lockheed Institute for Space Research, Stanford University, HEPL, 466 Via Ortega, Stanford, CA 94305, USA}
\altaffiltext{4}{Lockheed Martin Solar and Astrophysics Laboratory, 3251 Hanover Street, Palo Alto, CA 94304, USA}
\altaffiltext{5}{W. W. Hansen Experimental Physics Laboratory, Stanford University, Stanford, CA 94305, USA}

\begin{abstract}
Solar flares involve impulsive energy release, which results in enhanced radiation in a broad spectral and at a wide height range. In particular, line emission from the chromosphere can provide critical diagnostics of plasma heating processes. Thus, a direct comparison between high-resolution spectroscopic observations and advanced numerical modeling results can be extremely valuable, but has not been attempted so far.  
We present in this paper such a self-consistent investigation of an M3.0 flare observed by the Dunn Solar Telescope's (DST) {\it Interferometric Bi-dimensional Spectrometer} ({\it IBIS}) on 2011 September 24 that we have modeled with the radiative hydrodynamic code RADYN. We obtained images and spectra of the flaring region with IBIS in H$\alpha$~6563~\AA\ and \ion{Ca}{2} 8542~\AA, and with the {\it Reuven Ramaty High Energy Solar Spectroscope Imager} ({\it RHESSI}) in X-rays. The latter was used to infer the non-thermal electron population, which was passed to RADYN to simulate the atmospheric response to electron collisional heating. 
We then synthesized spectral lines and compared their shapes and intensities with those observed by IBIS and found a general agreement. In particular, the synthetic \ion{Ca}{2} 8542~\AA\ profile fits well to the observed profile, while the synthetic H$\alpha$~profile is fainter in the core than the observation. 
This indicates that H$\alpha$~emission is more responsive to the non-thermal electron flux than the \ion{Ca}{2} 8542~\AA\ emission. We suggest that a refinement of the energy input and other processes is necessary to resolve this discrepancy.
\end{abstract}


   \section{Introduction}
Energy release (e.g., by magnetic reconnection) in solar flares in general results in particle acceleration, plasma heating, and plasma wave (or turbulence) generation. Most of the released energy is transported by the accelerated particles  downward along magnetic field lines and deposited in the dense chromosphere by Coulomb collisions with ambient plasma in the so-called thick target model \citep[e.g.,][]{1971SoPh...18..489B, 1973ApJ...186..291P, 1976SoPh...50..153L}. Some energy may be transported by thermal conduction from directly heated coronal plasma \citep[e.g.,][]{1988ApJ...329..456Z, 2009A&A...498..891B}, and possibly by plasma waves \citep[e.g.,][]{2008ApJ...675.1645F, 2009ApJ...707..903H}. Observational signatures of the energy deposition include radiation in X-rays via bremsstrahlung of the electrons, gamma-rays via interaction of accelerated ions with background ions, various lines (e.g., H$\alpha$) and continuum emission from the heated plasma. These observations can provide useful diagnostics and help constrain mechanisms of energy release and particle acceleration, a fundamental question for solar flares.

Strong chromospheric lines, such as H$\alpha$~and \ion{Ca}{2} 8542~\AA\ are formed under conditions of non-local thermal equilibrium (non-LTE) and represent the response of the lower atmosphere to flare heating. Understanding the line formation is crucial for the correct interpretation of the observations and evolution of line intensities and profiles. Spectroscopic observations of such emission are primarily obtained from ground-based facilities, which have several advantages over space telescopes. Among these are the flexibility of real-time adjustments of pointing and exposure times, and wide ranges of available observing modes and filters. However, because of the limited field-of-view (FOV) (as a trade-off for high spatial resolution), our limited capability in predicting flares and seeing and our limited weather conditions, such ground-based spectroscopic observations of flares are rare \citep[e.g.][]{2012ApJ...748..138K, 2013ApJ...769..112D, 2012A&A...547A..34F} and thus very valuable.

Numerical modeling of flare line emission is a necessary step to interpret observational data and subsequently, to constrain flare mechanisms. Early modeling of atmospheric line emission were based on empirical flaring atmosphere models \citep[e.g.,][]{1984ApJ...282..296C, FangC.e-beam-Halpa.CaIIK.1993A&A...274..917F} or on radiative transfer simulations of an electron-beam heated chromosphere \citep[e.g.,][]{FisherG1985ApJ...289..414F, GanW1990ApJ...358..328G, DingMD.line-prof-beam-heating.1998A&A...332..761D}. \citet{2001MNRAS.326..943D} and \citet{2007ASPC..368..387B} later studied how the precipitation of electrons from the corona affects the line profiles. \citet{2009A&A...499..923K} focused on subsecond scale variations and solved 1-D radiative hydrodynamics of a solar atmosphere subjected to a subsecond electron beam heating to study the H$\alpha$~line emission. More sophisticated later models involved non-LTE radiative hydrodynamic calculations on timescales up to several tens of seconds: \citet{2005ApJ...630..573A} injected a power-law electron beam at the apex of a loop and tracked the atmospheric response and how the H$\alpha$~line evolved in time in response to the flux of electrons. 

In this paper we present a self-consistent, detailed comparison of line profiles from high-resolution spectroscopic flare observations and advanced radiative transfer hydrodynamic simulations using observationally inferred electron spectra as inputs. Such a comprehensive investigation {\it has not been attempted in the past} and can offer new insights to flare dynamics. Specifically, we obtain H$\alpha$~and \ion{Ca}{2} 8542~\AA\ line profiles of an M-class flare from the Interferometric Bi-dimensional Spectrometer (IBIS; \citealt{2006SoPh..236..415C, 2008A&A...481..897R}) instrument at the Dunn Solar Telescope (DST). The non-thermal electron spectra were inferred from hard X-ray (HXR) observations of the same flare by the {\it Reuven Ramaty High Energy Solar Spectroscope Imager} \citep[{\it RHESSI}; ][]{2002SoPh..210....3L}, which are then used as inputs to the RADYN code \citep{1992ApJ...397L..59C, 1997ApJ...481..500C, 2005ApJ...630..573A} to perform radiative transfer hydrodynamic simulations. Line profiles are synthesized from the simulation results and compared to the IBIS observations, completing a full circle of the investigation.

This paper is organized as follows. We describe the relevant observations in Section~\ref{Sect:obs} and the RADYN code in Section~\ref{sect:radyn}. We then present the simulation results and comparisons with observations in Section~\ref{sect:results} and finally the conclusions and discussion in Section~\ref{sect:conclusions}.

   \section{Observations}\label{Sect:obs}
\begin{figure}[htb]
 \centering
 \epsscale{1.03}
  \plotone{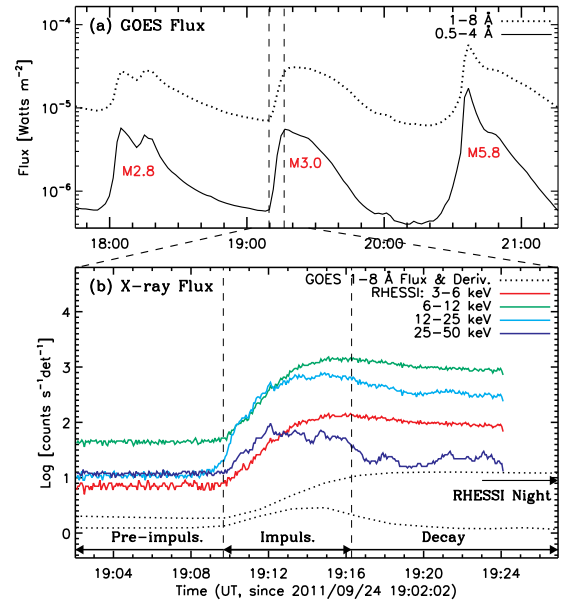}
  \caption{Temporal evolution of the flux measured on 24 September 2011. (a) {\it GOES} SXR flux measured every 3 seconds, showing the M3.0 flare and the preceding and succeeding M2.8 and M5.8 flares from the same AR; (b) {\it RHESSI} count rates in colored solid lines and {\it GOES} 1--8~\AA\ flux and its time derivative in black dotted lines, arbitrarily shifted vertically. 
The two vertical dashed lines indicate the impulsive phase.}
  \label{goes_rhessi_radyn}
\end{figure}

On 2011 September 24, an M3.0 class flare occurred in the NOAA active region (AR) 11302. As shown in Figure~\ref{goes_rhessi_radyn}, this flare started at 19:09~UT, reaching its maximum at 19:21 in the {\it GOES\ } 1--8~\AA\ flux and ending at 19:41~UT. {\it RHESSI\ } detected HXRs from the beginning of the impulsive phase (19:08) up to 19:24~UT. DST/IBIS observed footpoint emission from the chromosphere in H$\alpha$~and \ion{Ca}{2} 8542~\AA\ from 19:18 to 19:35~UT. The Atmospheric Imaging Assembly \citep[AIA; ][]{2012SoPh..275...17L} onboard the {Solar Dynamics Observatory} ({\it SDO}) observed the flare in (extreme) ultra-violet (UV); the 1700~\AA\ images of the flare ribbons were used to estimate the cross-sectional area of the flare loop.

  \subsection{Chromospheric Line Observations by IBIS}\label{sect:ibis}
IBIS is a dual Fabry-Perot system, capable of full-Stokes dual beam polarimetry of different spectral lines. The spectral lines and six polarization states are scanned sequentially and reconstructed into images of the Stokes vector during the data reduction process. During our observations, we scanned the chromospheric H$\alpha$~6563~\AA\ and \ion{Ca}{2} 8542~\AA\ lines and the photospheric \ion{Fe}{1} 6302~\AA\ line. Each line scan (with all polarization states and wavelength points) took about 25 seconds, giving an overall cadence of 95 seconds (including overhead such as filter wheel motions).  

The IBIS data reduction was performed with our graphical user interface (GUI) and included corrections for the dark and gain, the alignment of all channels, the wavelength shift due to the collimated Fabry Perot mount and the prefilter transmission profile. Speckle-reconstructed images from a simultaneous broadband (white-light) camera were used to destretch all spectral channels to remove the variations due to seeing. A polarimetric calibration was also performed, but is not relevant for our study, because we only analyze the intensity profiles. The flare occurred relatively late during the observing day, by which time the atmosphere around the telescope is usually heated up and the seeing is not ideal. Depending on the spectral line, only three to four of the ten scans had stable and favorable seeing to be used for this paper.

The line profiles were averaged along the ribbon. We estimated the area of the footpoints with pixels whose values are above 2500 DN, which is equal to a flux of 31250 DN s$^{-1}$. This corresponds to $\approx$ 50\% of the maximum intensity. Contours of this intensity level are illustrated in Figure~\ref{ibis_rastering}.
 
  \subsubsection{H$\alpha$~Observations}
The H$\alpha$~line is one of the most commonly observed lines in flares because of its high contrast and thus provides useful diagnostics. However, its interpretation is complex because the emission originates from a broad range of heights, from the upper photosphere to the lower chromosphere, and it is sensitive to the flux of non-thermal electrons precipitating to the chromosphere \citep{2009A&A...499..923K}.

There have been only a few published observations which show the temporal evolution of the H$\alpha$~profiles from the start of a flare because of the difficulty in setting and holding the spectrograph slit on a flare footpoint \citep[e.g.][]{1961BAICz..12...47K, 1985ApJ...288..353C, 2007A&A...461..303R, 
2011A&A...535A.123R, 2013ApJ...769..112D}. With IBIS we obtained three H$\alpha$~scans with good seeing, starting after the impulsive phase, at 19:22:40 UT, 19:24:15 UT and 19:32:09 UT; each of them lasted 27 seconds and included 24 wavelength points in a wavelength range of $\pm$ 2~\AA\ from the line center.

  \subsubsection{\ion{Ca}{2} 8542~\AA\ Observations}
The \ion{Ca}{2} infrared triplet ($\lambda$=8498, 8542 and 8662~\AA) provides a useful diagnostic of the solar chromosphere because these lines result from the transition between the upper 4p $^2$P$_{1/2,3/2}$ levels and the lower metastable 3d $^2$D$_{3/2,5/2}$ levels. (There are no allowed electric dipole transitions to the ground state.).

The \ion{Ca}{2} 8542~\AA\ line is a natural tracer of solar activity, having a dual nature: its outer wings sample the solar photosphere and the intensity in the far wings is very sensitive to the presence of magnetic structures \citep[c.f.][]{2006A&A...452L..15L}, while the core is affected by chromospheric motions. Despite this and the fact that in the near infrared wavelengths the terrestrial atmospheric turbulence is reduced, the \ion{Ca}{2} infrared triplet observations are still scarce and only possible with ground-based instruments. In this paper we use imaging spectroscopy in the \ion{Ca}{2} 8542~\AA\ line. IBIS scanned the line within 25 seconds, including 23 wavelength points  along a spectral range of $\pm$ 2.3~\AA\ from the line center. The observation times with good-seeing started at 19:22:15 UT, 19:23:49 UT, 19:26:59 UT, and 19:30:08 UT. 

\begin{figure*}[htb]
 \centering
  \epsscale{1.2}
  \plotone{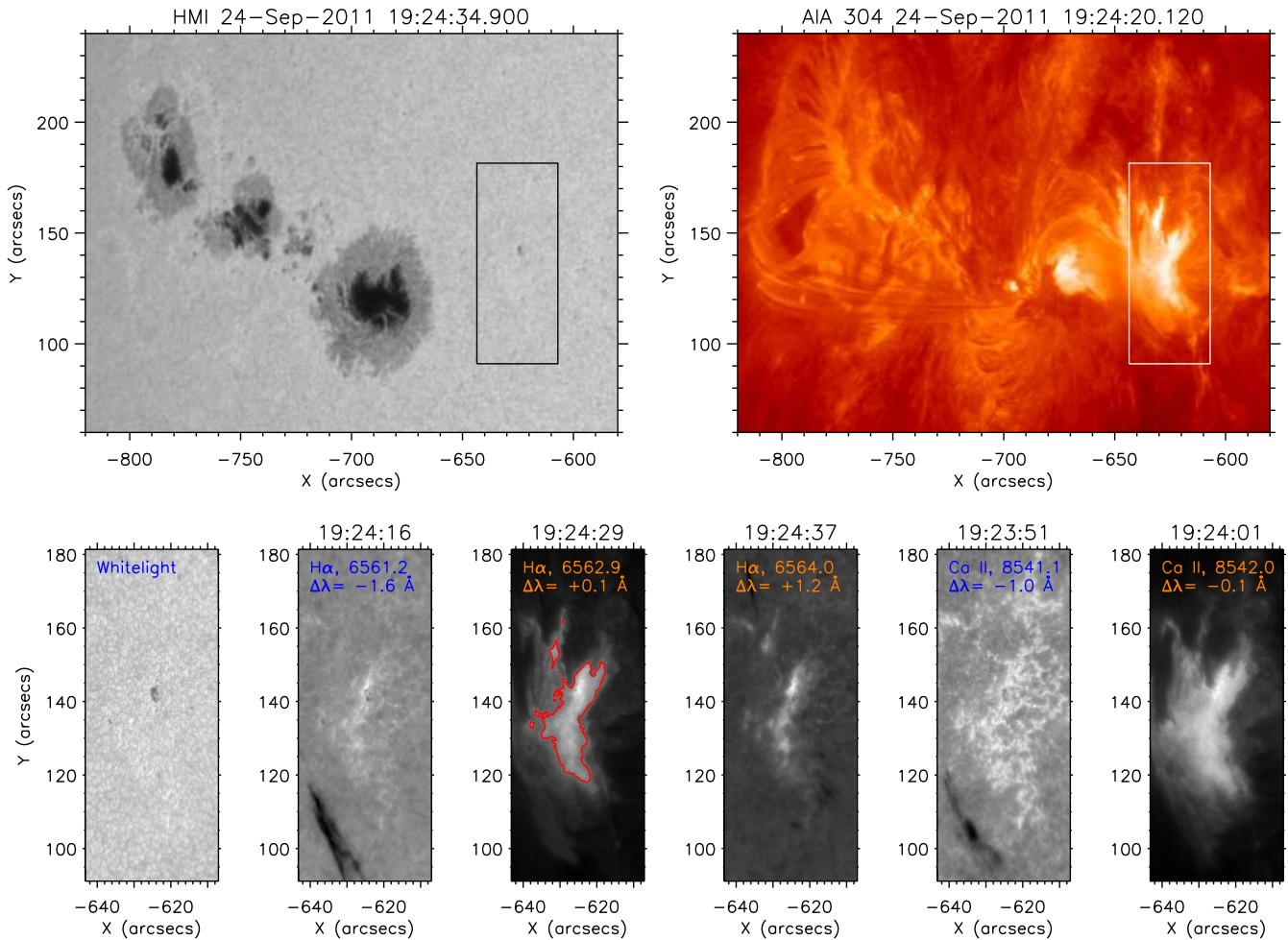}
  \caption{Overview of AR 11302 seen by {\it SDO} (top row) Helioseismic and Magnetic Imager (HMI) in continuum and AIA in 304~\AA, and by IBIS (bottom row) in white light, H$\alpha$~(three wavelengths) and \ion{Ca}{2} 8542~\AA\ (two wavelengths). The box in the {\it SDO} images denotes the FOV of IBIS, which covers the western flare ribbon. The red contours in the 6562.9~\AA\ image show the area that was averaged to obtain an average intensity profile of the ribbon.}
  \label{ibis_rastering}
\end{figure*}

Figure~\ref{ibis_rastering} shows two snapshots from {\it SDO} of AR 11302 at 19:24 UT (top). The box denotes the FOV of IBIS, which was centered on a small pore west of the active region and covered one of the flare ribbons. The bottom row shows example images from IBIS with their wavelengths labeled. While the blue wing images (6561.2~\AA\ and 8541.1~\AA) show a filament in the lower left corner and a brightening due to the ribbon emission, the images near the line cores (6562.9~\AA\ and 8542.0~\AA) resemble the emission of AIA~304. During our observations, the ribbon is seen to expand and the filament becomes smaller and weaker in intensity.

  \subsection{\textbf{\textit{RHESSI}} Observations}\label{sect:rhessi}
{\it RHESSI} had full coverage of the impulsive phase of the flare and part of the decay phase until its sunset, which started at 19:23:52~UT (see Figure~\ref{goes_rhessi_radyn}).

\begin{figure*}[htb]
 \centering
  \epsscale{1.2}
  \plotone{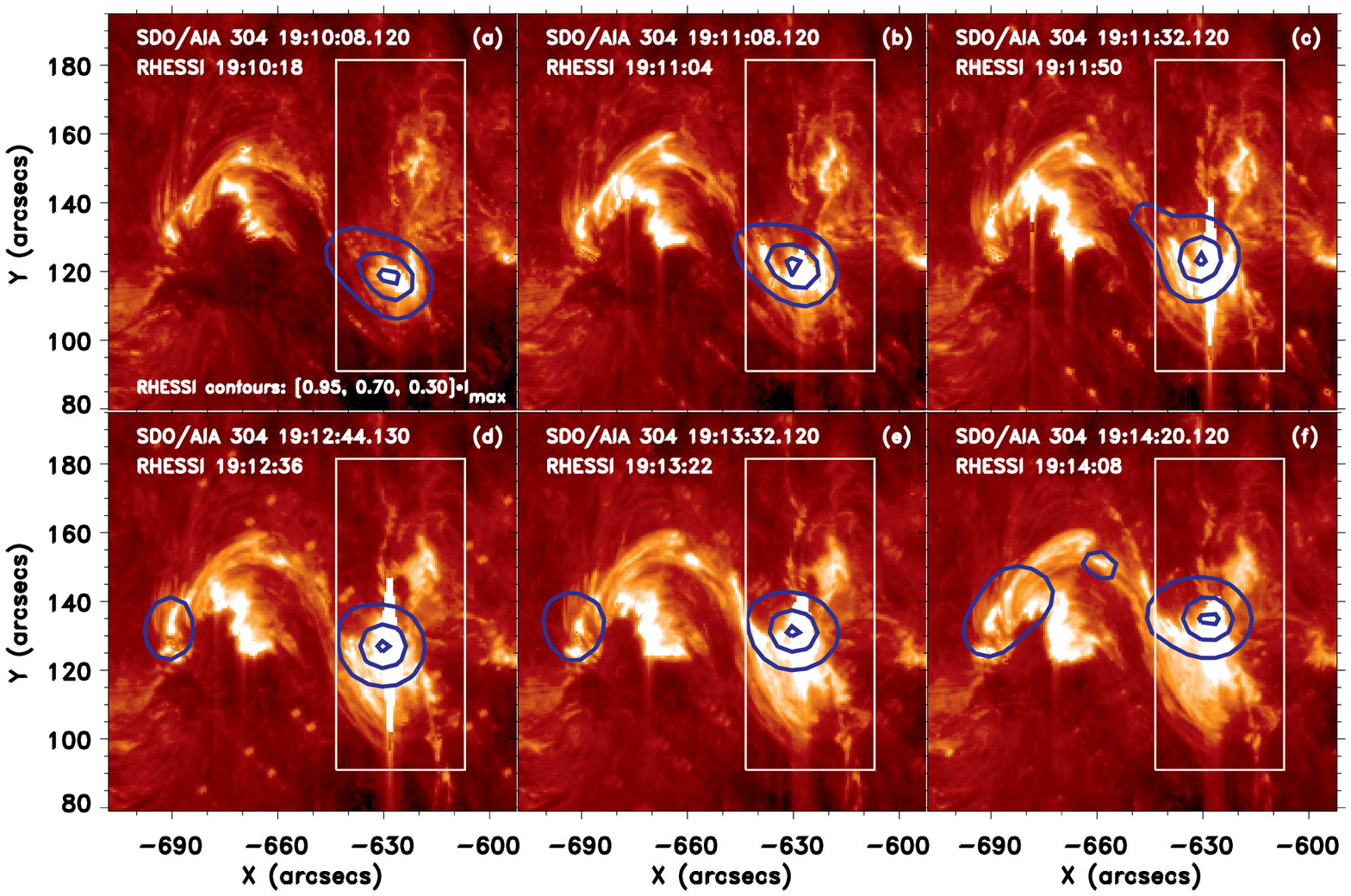}
  \caption{{\it RHESSI} 25-50 keV emission contours over {\it SDO}/AIA 304~\AA\ images, during the impulsive phase of the flare, from 19:10:18 to 19:14:20. The contours represent the 95, 70 and 30\% of the maximum intensity at each time interval. The white box represents the FOV of IBIS.}
  \label{rhessi_sdo}
\end{figure*}

Using the CLEAN algorithm, we reconstructed images at 25-50 keV integrated over 52~s intervals from 19:08:52 to 19:35:44. These images revealed two HXR footpoint sources and a loop top source at later times (see Figure~\ref{rhessi_sdo}). The western footpoint is brighter and moves faster than the eastern one, which is located in a stronger magnetic field. Such asymmetric footpoint emission is understood as a result of asymmetric magnetic mirroring \citep[e.g.,][]{WangH1995ApJ...453..505W, 2007A&A...471..705J, 2009ApJ...693..847L, 2012ApJ...756...42Y}.

  \subsubsection{Inferring the non-thermal electron distribution}\label{sect:rhessi_spectra}
We can infer the non-thermal electron distribution and thus its energy flux by analyzing {\it RHESSI} X-ray spectra. The moderate count rates of this M3.0 flare allow imaging spectroscopy of spatially resolved sources only during the HXR peak of the impulsive phase. To cover the temporal evolution of the flare, we thus chose to fit the spatially integrated spectra, which are dominated by the western HXR footpoint (see Figure~\ref{rhessi_sdo}) that coincides with the flare kernel position observed by IBIS. This provides reasonable diagnostics for the electron energy flux as an input to our RADYN simulation of that kernel.

Because each of {\it RHESSI}'s nine germanium detectors has slightly different characteristics and makes independent measurements, we analyzed the spectra of individual detectors separately. We then used the means and standard deviations of the fitting parameters to obtain the best-fit parameter and uncertainties. This has several advantages (e.g., avoiding energy smearing) over the conventional approach of directly fitting the average count spectra of all detectors \citep[for details, see,][]{2008ApJ...676..704L, 2009ApJ...699..968M}.
We excluded detectors~2 and 7 because of their abnormally high threshold and/or low energy resolution \citep{2002SoPh..210...33S}. We also excluded detector~6 due to its consistently higher $\chi ^2$ values of the fitting results than those of the other detectors. For each of the remaining six detectors, we obtained photon spectra by integrating 30~s intervals from 19:09:30 to 19:13:00~UT and 60~s intervals from 19:13:00 to 19:23:52~UT until the spacecraft night.

Using the standard Object Spectral Executive (OSPEX) software package \citep{2006ApJ...643..523B}, we applied corrections for albedo, instrumental emission lines, pulse pileup and the detector response matrix (DRM). We fitted the spectra with a thermal component plus a thick-target, non-thermal component consisting of a broken power law, 
\begin{equation}
F(E) = (\delta -1)\frac{\mathcal{F}_c}{E_c} \Big( \frac{E}{E_c} \Big)^{-\delta} \,
\label{power-law.eq}
\end{equation}
for energies $E>E_c$ and a power law with a fixed index $\delta=-2$ for $E<E_c$. Here
$\mathcal{F}_c=\int^{\infty}_{E_c} F(E) \;dE$ is the total electron flux above $E_c$.

\begin{figure}[!bth]
\centering
  \epsscale{1.1}
  \plotone{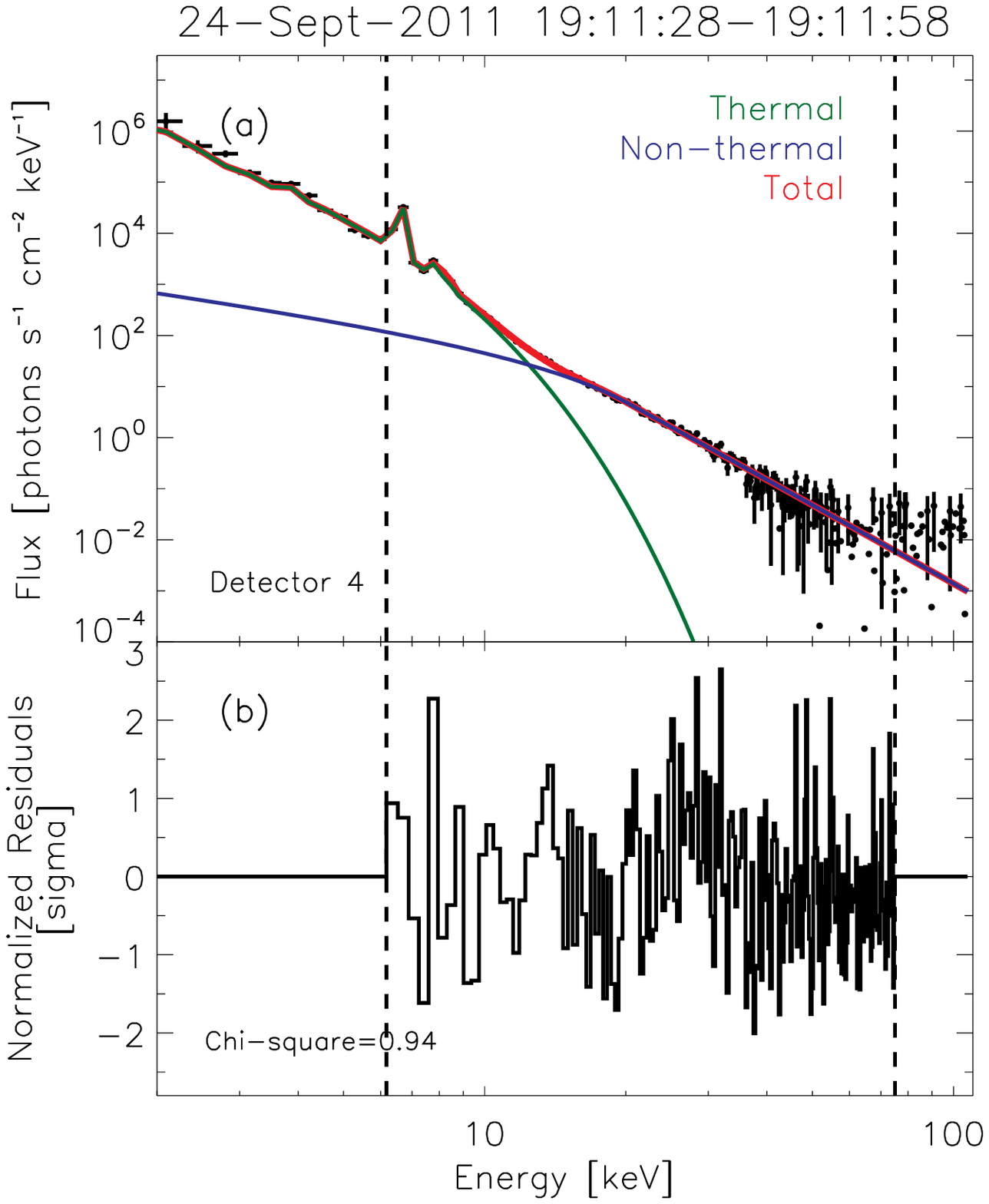}
  \caption{(a) {\it RHESSI} photon spectrum from detector 4 taken at 19:11:28-19:11:58, during the impulsive phase. The green and blue lines represent the fitted thermal and non-thermal power-law components respectively. Other contributions to the spectrum are not shown. The red line is the final fit, taking into account all the components. The two vertical dotted lines mark the energy range over which the spectral fit was performed. (b) Residuals of the fit normalized to 1$\sigma$ at each energy.}
     \label{rhessi_spectra_time}
\end{figure}

Figure~\ref{rhessi_spectra_time}(a) shows a spectrum of detector~4, as an example, during the impulsive phase, from which we can clearly see the dominance by the thermal component at low energies and by the non-thermal component at high energies. The green and blue lines show the fitted thermal and non-thermal components, respectively, and the red line is the total fit, summing all the components.

\begin{figure}[!htb]
\centering
  \epsscale{1.1}
  \plotone{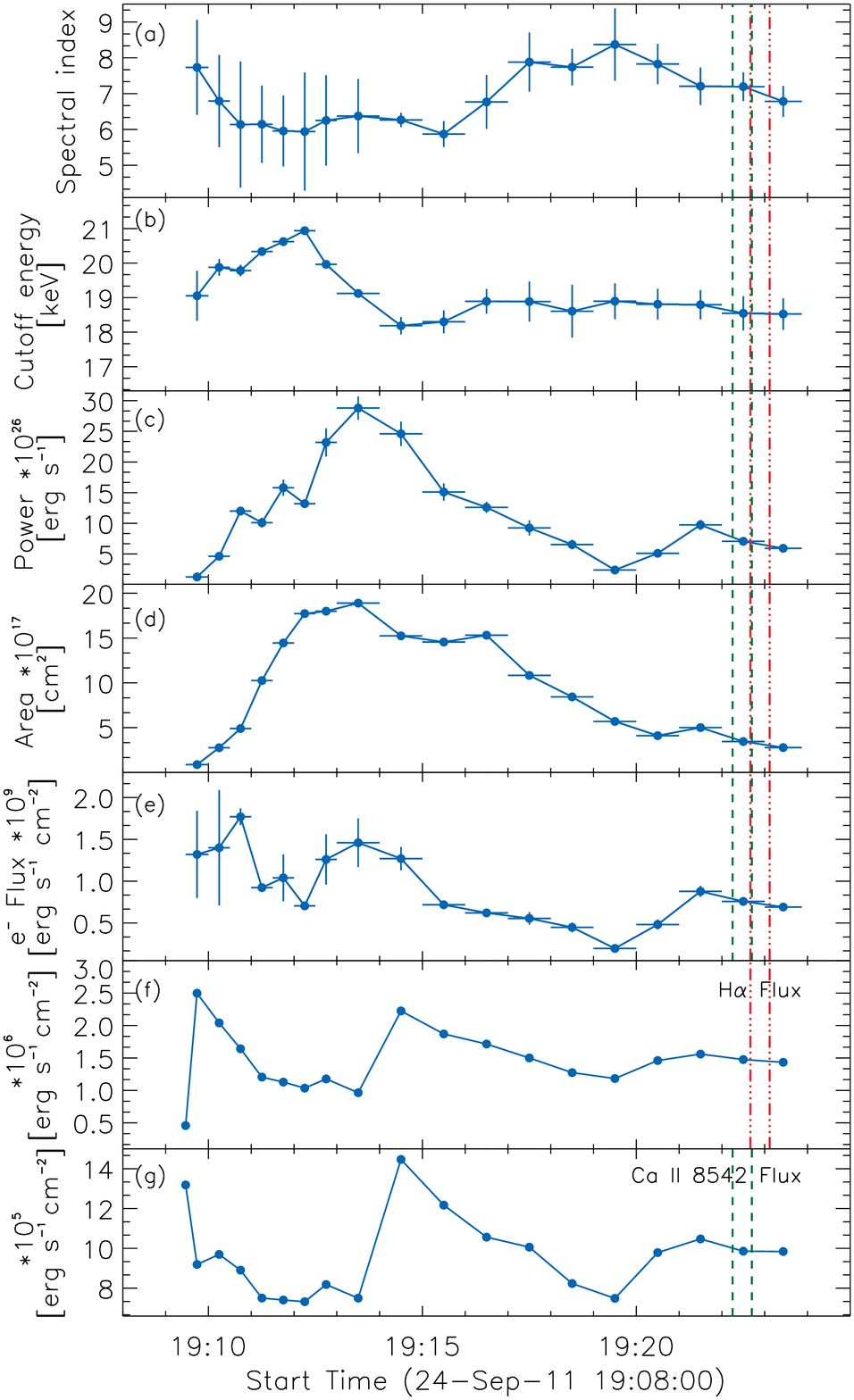}
  \caption{Variation of the non-thermal electron spectral parameters obtained from fitting the {\it RHESSI} data. (a) Mean spectral index; (b) Low cutoff energy; (c) Non-thermal power; (d) Area of the AIA 1700~\AA\ footpoints; (e) Electron energy flux obtained by dividing the power by the footpoints area. (f) H$\alpha$~synthetic light curve obtained with RADYN ($\Delta \lambda$=3.8~\AA); (g) \ion{Ca}{2} 8542~\AA\ synthetic light curve obtained with RADYN ($\Delta \lambda$=4~\AA) [See text]. The error bars in the $x$-axis represent the integration time of the spectrum (panels (a)-(e)) and in the $y$-axis the standard deviation resulting from the combination of the 6 {\it RHESSI} detectors. In panels (f) and (g), flux values have only been plotted at the integration interval for clarity. The red and green vertical lines represent the time range during which the H$\alpha$~and \ion{Ca}{2} 8542~\AA\ line profiles shown in this paper were obtained.}
  \label{fig_parameters_time}
\end{figure}

The temporal evolution of the power-law index $\delta$ and energy cutoff $E_c$ of the non-thermal component are shown in Figures~\ref{fig_parameters_time}(a) and \ref{fig_parameters_time}(b). We find a soft-hard-soft spectral evolution, with the hardest spectra (lowest $\delta$) occurring at 19:12--19:15~UT.

\begin{figure*}[!tb]
  \epsscale{1.2}
  \plotone{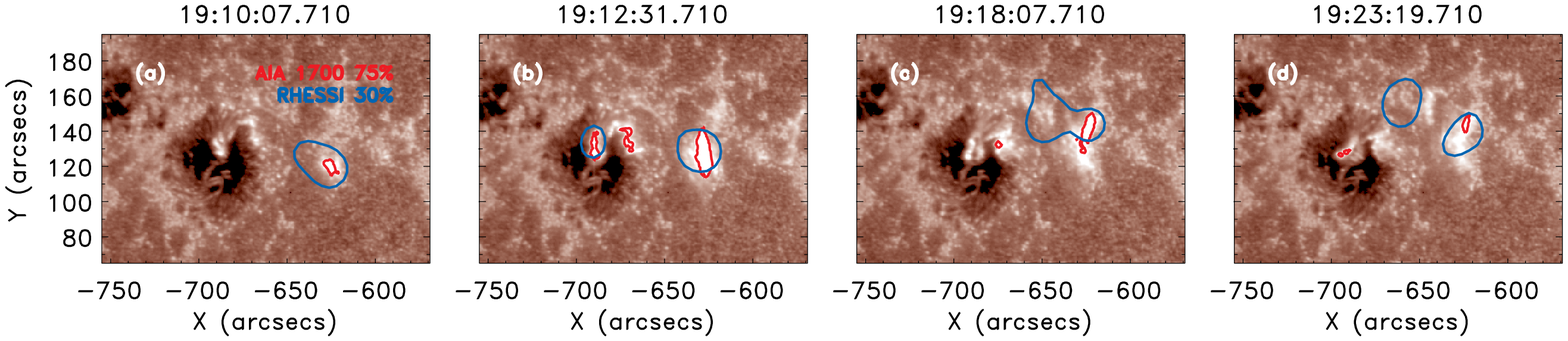}
  \caption{{\it SDO}/AIA 1700~\AA\ images showing the evolution of the flare ribbons. The red contours represent the footpoints areas having 75\% of the maximum intensity and the blue contours, the 30\% of the {\it RHESSI} 25-50 keV maximum flux}.
  \label{ribbons1700}
\end{figure*}

From {\it RHESSI} we obtain the total rate of injection of electrons. To estimate the flux of electrons (or energy flux ${\cal E}(E)=E\times F(E)$) within the loop, it requires the knowledge of its cross-sectional area. The spatial resolution ($\geq 7\arcsec$) of the combined {\it RHESSI} detectors 3--9 used for HXR imaging is insufficient for resolving the flare footpoints. We thus approximated this area with {\it SDO}/AIA 1700~\AA\ ($1 \farcs 2$ resolution) kernels within the blue contours at 75\% of the maximum intensity, as shown in Figure~\ref{ribbons1700}. This approximation is justified by the fact that the primary 1700~\AA\ kernel of the largest area and highest intensity is cospatial with the {\it RHESSI} HXR footpoint, as shown in Figure~\ref{ribbons1700}. Similar cospatiality between H$\alpha$~kernels and HXR footpoints has also been reported \citep[e.g.][]{2007ApJ...658L.127L}. However, we also notice some small 1700~\AA\ kernels without a corresponding HXR source, which could be caused in part by {\it RHESSI}'s limited dynamic range of the order of 1:10.

As the flare evolves, the contours associated with the AIA 1700~\AA\ easternmost footpoint are not anymore cospatially aligned with the {\it RHESSI} 25-50 keV emission. The western footpoint (observed by IBIS) is the main contributor to the HXR flux and still coincides with the 1700~\AA\ emission. In addition, we also see HXR emission at different locations than 1700~\AA\ emission [see e.g. Figure~\ref{ribbons1700}(c) and (d)]. One possibility is that such HXRs are emitted from the loop-top source [see Figure~\ref{rhessi_sdo}(f)]. Nevertheless, this mismatch could result in an overestimate of the flare loop cross-sectional area and thus underestimate of the electron energy flux. We estimated that the uncertainty in the inferred area is about 16\%.

Images taken between 19:11 and 19:13~UT contain several saturated pixels, which we discarded by imposing an upper threshold of 16000~DN. The resulting areas were then interpolated to the time intervals used for {\it RHESSI} spectral fitting and are shown in Figure~\ref{fig_parameters_time}(d). Finally, by dividing the electron power by the estimated loop cross-sectional area, we obtained electron energy flux, whose temporal evolution is shown in Figure~\ref{fig_parameters_time}(e). 

  \section{RADYN simulations}\label{sect:radyn}
We used the RADYN code of \citet{1997ApJ...481..500C}, including the modifications of \citet{1999ApJ...521..906A} and \citet{2005ApJ...630..573A}, to simulate the radiative-hydrodynamic response of the lower atmosphere to energy deposition by non-thermal electrons in a single flare loop.  

  \subsection{General Description of the RADYN Code}
The code solves simultaneously the equations of hydrodynamics, population conservation, and radiative transfer implicitly on a one-dimensional adaptive grid \citep{1987JCoPh..69..175D}, as described by \citet{1992ApJ...397L..59C}. 

For radiative transfer calculations, atoms important to the chromospheric energy balance are treated in non-LTE. These include six-level plus continuum hydrogen, six-level plus continuum, singly ionized calcium, nine-level plus continuum helium, and four-level plus continuum, singly ionized magnesium. Line transitions treated in detail are listed in Table~1 of \citet{1999ApJ...521..906A}. Complete redistribution is assumed for all lines, except for the Lyman transitions in which partial frequency re-distribution is mimicked by truncating the profiles at 10 Doppler widths \citep{1973ApJ...185..709M}. Other atomic species are included in the calculation as background continua in LTE, using the Uppsalla opacity package of \citet{Gustafsson}.

Additions of hydrodynamic effects due to gravity, thermal conduction, and compressional viscosity to the original RADYN code were described by \citet{1999ApJ...521..906A}. Later additions by \citet{2005ApJ...630..573A} included photoionization heating by high-temperature, soft X-ray emitting plasma, optically-thin cooling due to thermal bremsstrahlung and collisionally excited metal transitions, and conductive flux limits to avoid unphysical values in the transition region of large temperature gradients. 

  \subsection{Simulation Setup}
We assumed a single quarter circle loop geometry in a plane-parallel model atmosphere, discretized in 191 grid points. The model loop is 10~Mm in height. We assume a symmetric boundary condition at the loop apex ($z=10$~Mm). We note from the observations (Figure~ \ref{ribbons1700}) that the footpoints appear to move during the flare but not more than its diameter so that our assumption of a single loop is a reasonable approximation. 

\begin{figure}[htb]
  \epsscale{1.0}
  \plotone{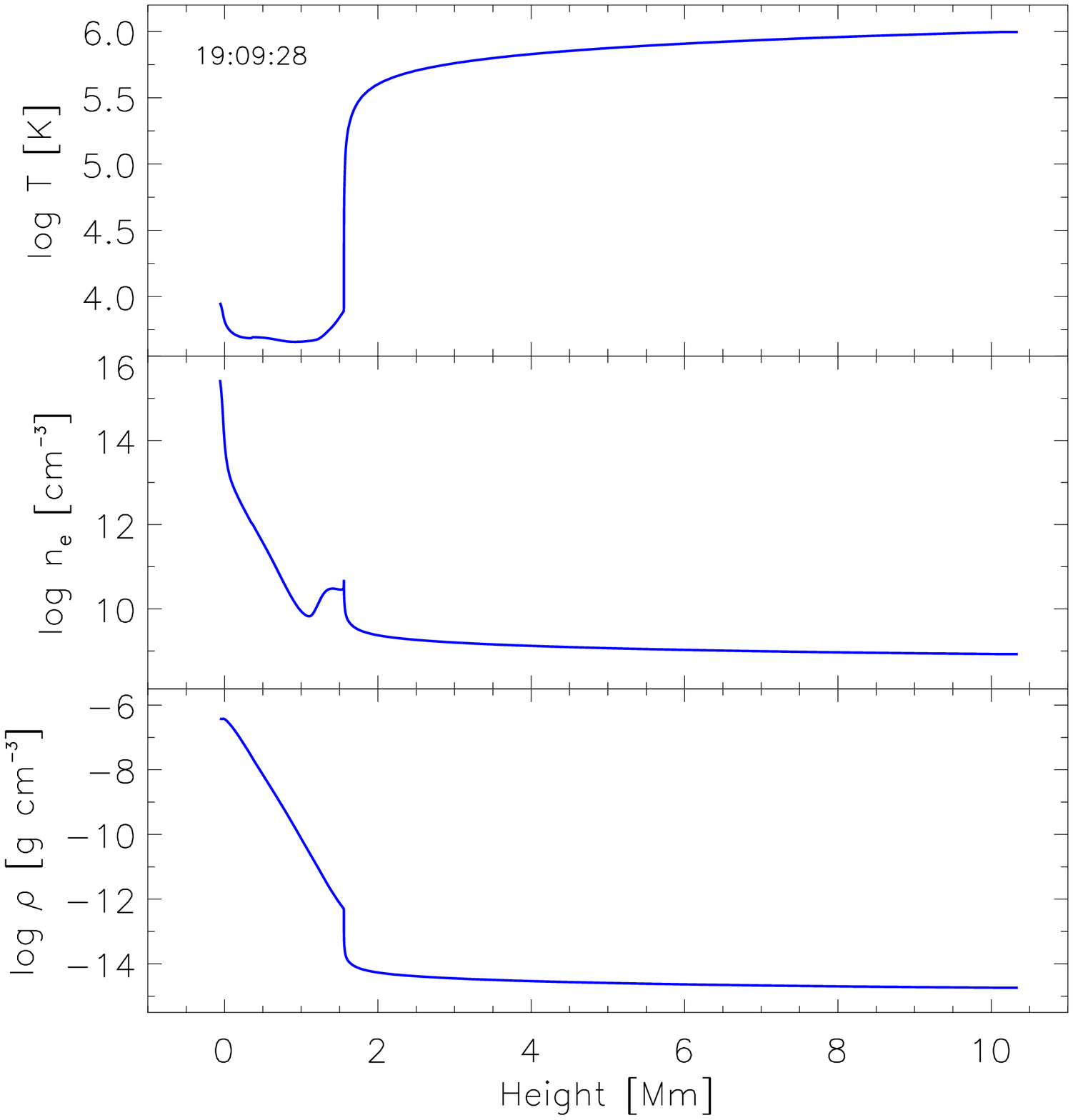}
  \caption{Initial atmosphere FP2, from \citet{1999ApJ...521..906A}, in hydrodynamic equilibrium.}
  \label{preflare}
\end{figure}

The initial atmosphere (Figure~\ref{preflare}) was adopted from the FP2 model of \citet{1999ApJ...521..906A}, which is generated by adding a transition region and corona to the model atmosphere of \citet{1997ApJ...481..500C}. The temperature was fixed at 10$^6$~K at the loop top and no external heating was provided. This allowed the the atmosphere to relax to a hydrodynamic equilibrium state.

Initially the bottom boundary is located in the upper photosphere; the chromosphere is at 0.9~Mm from the bottom of the loop and the transition region is at a distance of 1.56~Mm from the bottom. During the evolution of the atmosphere we assume open boundaries at the bottom and apex of the loop, extrapolating if necessary. 

The non-thermal electron heating was calculated from the power-law spectrum provided by {\it RHESSI} spectral fits (see Figure~\ref{fig_parameters_time}). It was included in RADYN as a source of external heating in the equation of internal energy conservation ($Q$ term of equation~(3) in \citet{1999ApJ...521..906A}) and has been updated every integration time interval based on {\it RHESSI} data. Intermediate times have been interpolated.

   \subsection{Chromospheric Response to Non-Thermal Electrons}\label{sect:radyn_chrmsph}
\begin{figure}[htb]
\centering
  \epsscale{1.0}
  \plotone{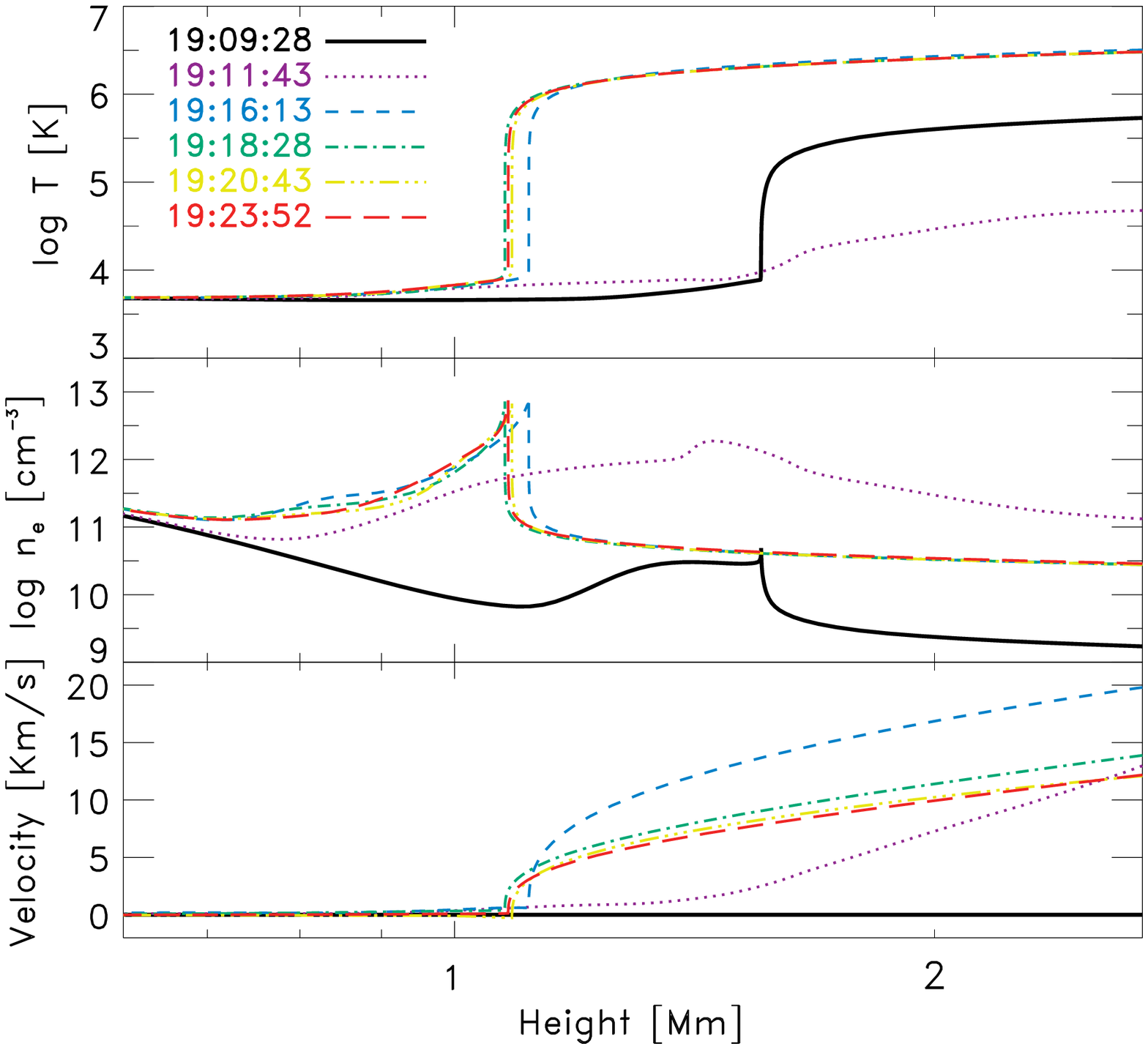}
  \caption{Chromospheric spatial distribution of the temperature, electron density and velocity (at different time steps). Positive velocities correspond to upflowing plasma and negative velocities to downflowing plasma. The back solid line is the initial atmosphere in hydrodynamic equilibrium. The X-axis is in logarithmic scale.}
  \label{evol_chromosphere}
\end{figure}

In general, the hydrodynamic evolution of the atmosphere is qualitatively similar to the F09 case reported by \citet{1999ApJ...521..906A}. Here we focus on a narrow region within a distance range of $z=$0.62--2.7~Mm, as shown in Figure~\ref{evol_chromosphere}, which covers the upper chromosphere and transition region. This is the region where the dynamic evolution relevant to the H$\alpha$~and \ion{Ca}{2} 8542~\AA\ line emission occurs.

The electrons injected at the loop apex travel downwards, losing energy by Coulomb collision, and heat up the chromospheric plasma. This leads to an overpressure which drives chromospheric evaporation with a plasma upflow starting at about 5 seconds (19:09:33). This also causes the transition region to move upwards (see e.g. violet dotted line at 19:11:43 in Figure~\ref{evol_chromosphere}). 

As time proceeds and the injected non-thermal electron flux decreases, the atmosphere relaxes (see Figure~\ref{fig_parameters_time}). The temperature in the corona and the position of the transition region are directly related to the flux of injected electrons. This explains the small displacements of the transition region that can be seen over time in Figure~\ref{evol_chromosphere} in the lines illustrating the spatial distribution at 19:18:28 (green), 19:20:43 (yellow) and 19:23:52 (red).

 \subsection{Formation and Evolution of Spectral Lines}
In order to compare the synthetic H$\alpha$~and \ion{Ca}{2} 8542~\AA\ profiles with the observations, we integrate the model intensities at each wavelength for 25 and 27 seconds respectively, which is the cadence of the observed profiles. A constant microturbulent velocity of 4.5~km~s$^{-1}$ has been applied to compensate for the lack of small scale random motions in the model \citep{2012A&A...543A..34D}.

Since the atmosphere at the beginning of the run is in an equilibrium condition, we treat the line profile at this time as the quiet Sun profile, which is subtracted from the intensity profiles at other times, obtaining the so-called excess line profile \citep{1998A&A...337..294H, 2008PASJ...60...95M}, which will allow us to better interpret how the flux of electrons affects the flare emission. The observed H$\alpha$~and \ion{Ca}{2} 8542~\AA\ profiles have a wavelength range of 3.8~\AA\ and 4~\AA\, which is also used for displaying the simulated profiles.

In order to better understand how the atmospheric evolution affects the chromospheric emission, we write the formal solution of the transfer equation for emergent intensity \citep{1997ApJ...481..500C}:
\begin{equation}
I_{\nu}^{0} = \frac{1}{\mu} \int_{\tau_{\nu}} S_{\nu} e^{-\frac{\tau_{\nu}}{\mu}} d\tau_{\nu}=
\frac{1}{\mu} \int_{z} S_{\nu} \chi_{\nu} e^{-\frac{\tau_{\nu}}{\mu}} dz = \frac{1}{\mu} \int_{z} C_i dz \,\,,
\label{eq_contribution_function}
\end{equation}
where $\chi_{\nu}$ is the monochromatic opacity per unit volume; $S_{\nu}$ is the source function, which is defined as the ratio between the emissivity to the opacity of the atmosphere; $\tau_{\nu}$ is the monochromatic optical depth and the integrand $C_i$ is the so called intensity contribution function, which represents the emergent intensity emanating from height $z$.

  \subsubsection{Evolution of the H$\alpha$~Line Profile}

\begin{figure*}[htb]
  \epsscale{1.2}
  \plotone{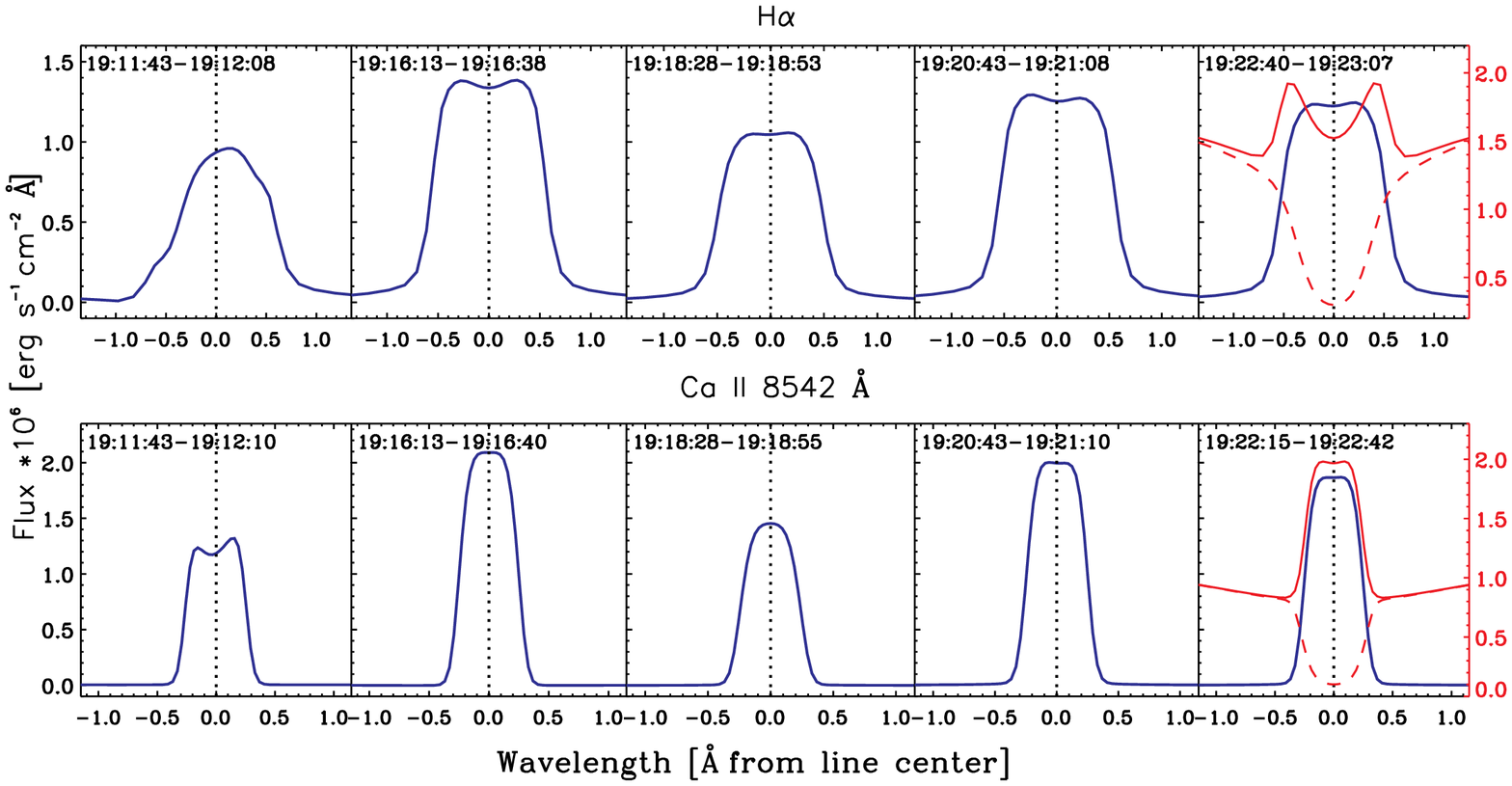}
  \caption{Time evolution of the averaged H$\alpha$~(top row) and \ion{Ca}{2} 8542~\AA\ (bottom row) excess line profiles, where the quiet Sun emission (red dashed line in the rightmost column) has been subtracted, integrated over every 25 and 27 seconds respectively. The dotted vertical line indicates the line center at rest. The red solid line in the rightmost column shows the line profiles including the quiet Sun emission, which diminishes when the quiet Sun is subtracted.}
  \label{evol_ha_ca_lines_radyn}
\end{figure*}

The top row of Figure~\ref{evol_ha_ca_lines_radyn} shows the synthesized H$\alpha$~excess profiles, which are asymmetric early during the flare and become almost symmetric later when the atmosphere relaxes. The line profile including the quiet Sun emission (red solid line in the right column of Figure~\ref{evol_ha_ca_lines_radyn}) shows a dip in the line core.

By integrating the intensity along the line profile, and subtracting the quiet Sun emission, we estimated the evolution of the H$\alpha$~excess flux in time. Figure~\ref{fig_parameters_time}(f) shows the H$\alpha$~light curve, where flux has been averaged during the integration time of each {\it RHESSI} spectrum. As the plasma is pushed upwards, the chromospheric evaporation takes place and the H$\alpha$~flux decreases. After 19:14:24 the atmosphere is more stable and the H$\alpha$~flux varies with the flux of the injected non-thermal electrons.

\begin{figure}[htb]
  \epsscale{1.0}
  \plotone{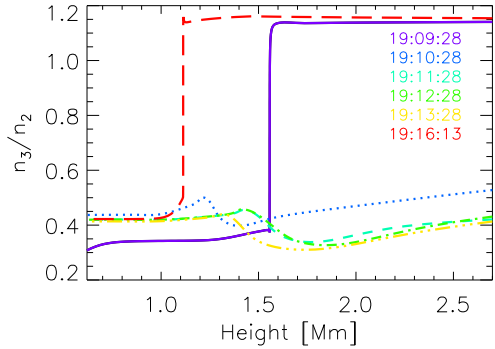}
  \caption{Height stratification of the ratio between the hydrogen energy level $n_3$ and the energy level $n_2$.}
  \label{n3_n2}
\end{figure}

Between 19:10 and 19:13, the density population at the energy level $n_3$ of the hydrogen atom decreases by almost a factor of two with respect to the density population at the energy level $n_2$ (see Figure~\ref{n3_n2}; therefore the ratio $n_3/n_2$ at this time range decreases. The fact that a H$\alpha$~photon is emitted by the transition from $n_3$ to $n_2$, that explains the decrease of the H$\alpha$~flux at these times. Calcium atoms have a similar behavior, which explains the similar decrease shown in Figure~\ref{fig_parameters_time}(g).

\begin{figure*}[htb]
  \epsscale{2.2}
  \plottwo{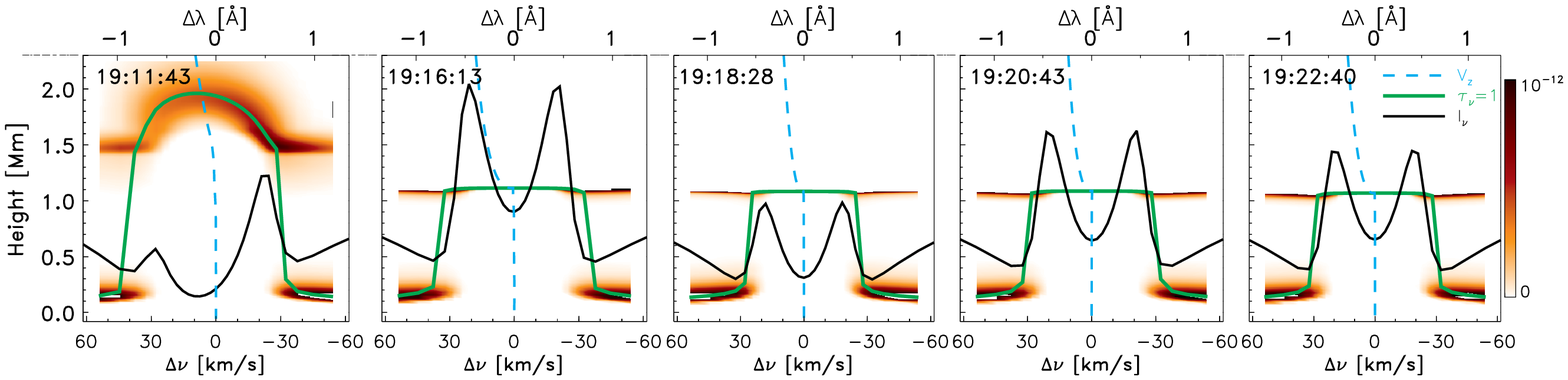}{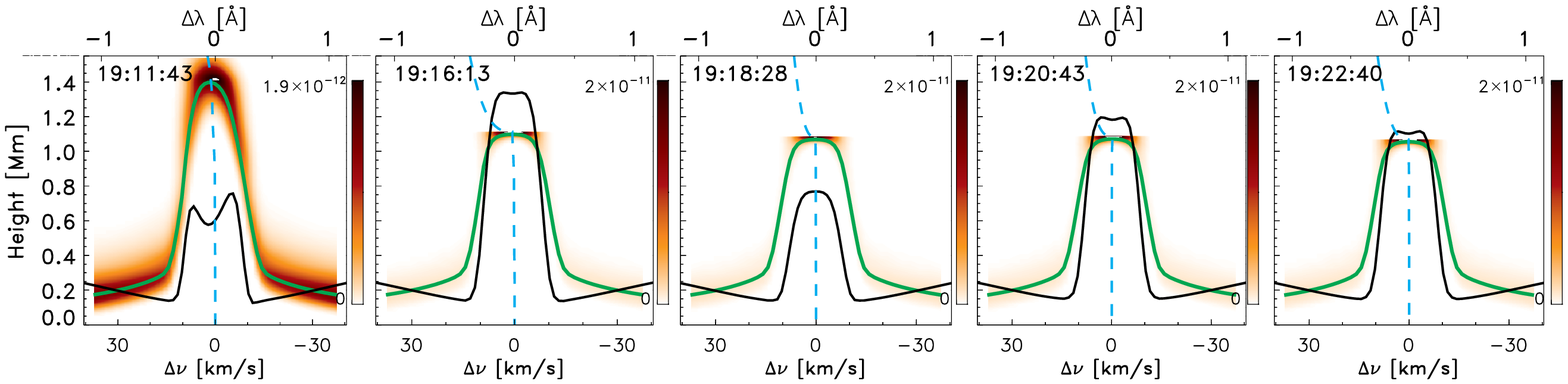}
  \caption{Intensity contribution function for H$\alpha$~(top) and \ion{Ca}{2} 8542~\AA\ (bottom) shown as reversed color maps (darker for higher values), as a function of frequency (bottom axis) and height (vertical axes). The frequencies are in velocity units such that positive and negative velocities represent upflows and down flows of plasma. The blue line represents the atmospheric velocity stratification; the green line the height at which $\tau_{\nu}=1$ and the black line, the line profile (including the quiet Sun emission).}
  \label{evol_ci_ha_ca}
\end{figure*}

The top row of Figure~\ref{evol_ci_ha_ca} shows the intensity contribution function, $C_i$ (increasing from bright to dark), for H$\alpha$~at the same times as in Figure~\ref{evol_ha_ca_lines_radyn}. The line frequencies are in velocity units, where positive velocities represent plasma moving upwards, towards the corona and negative velocities denote material moving downwards. The blue line represents the atmospheric velocity stratification and the black line, the line profile (including the quiet Sun emission, as the red profiles of Figure~\ref{evol_ha_ca_lines_radyn}). The green line represents the height at which $\tau_{\nu}=1$, showing us that the height formation of H$\alpha$~wings is coming from the lower chromosphere ($\approx$0.2~Mm) and is constant in time, while the height formation of the core varies in time during the two minutes, moving towards lower heights.

Studying how the contribution function changes in time (Figure~\ref{evol_ci_ha_ca} top) with height and wavelength, we found that the H$\alpha$~emission profile becomes broad and centrally reversed due not only to non-thermal effects when the atmosphere is bombarded by energetic electrons \citep{1984ApJ...282..296C}, but also to the temperature spatial distribution and the sudden behavior change of the source function in a very thin atmospheric layer.

  \subsubsection{Evolution of the \ion{Ca}{2} 8542~\AA\ Line Profile}
The synthesized \ion{Ca}{2} 8542~\AA\ excess profiles are shown at the bottom row of Figure~\ref{evol_ha_ca_lines_radyn}. The asymmetry in the line profile at early times is not as prominent as for H$\alpha$~and during the flare the line becomes almost symmetric, specially in the core.

To obtain the temporal evolution of the \ion{Ca}{2} 8542~\AA\ intensity, we integrated the intensity along the line profile with the quiet Sun subtracted (Figure~\ref{fig_parameters_time}(g)). The \ion{Ca}{2} 8542~\AA\ excess light curve follows a similar behavior than H$\alpha$, peaking both fluxes at the same time.

Following equation~\ref{eq_contribution_function}, the intensity contribution function $C_i$ for the \ion{Ca}{2} 8542~\AA\ line is represented in the bottom row of Figure~\ref{evol_ci_ha_ca}. We can see that $C_i$ becomes stronger in the core of the line and presents a symmetric behavior in the wings, being sensitive to plasma velocity changes. The monochromatic optical depth (green line) shows us that the formation of the wings is constant in time and located below 0.2~Mm; the height formation of the core moves towards lower heights for almost seven minutes. Afterwards it is stable at 1.05~Mm.

  \section{Comparing observed and synthetic line profiles}\label{sect:results}
{\it RHESSI} entered night at 19:23:52 UT, whereas the first available IBIS observation with decent seeing started at 19:22:UT. The two observations thus overlapped for $\approx$ 1 minute. In the following Section we will compare the synthetic profiles obtained from the radiative hydrodynamic code, using the {\it RHESSI} spectral information as input, with the line profiles observed by IBIS at 19:22:40 UT for H$\alpha$~and at 19:22:15 UT for \ion{Ca}{2} 8542~\AA.

  \subsection{Calibration of the quiet Sun profiles}\label{sect:calibrate_qs}
To be able to compare the observed line profiles with the synthetic profiles, we first have to calibrate the spectral lines of IBIS and RADYN to the same reference system. We use the continuum emission of the Fourier Transform Spectrometer (FTS) atlas taken at the McMath-Pierce Telescope \citep{brault_neckel} as reference. By doing so, both observed and synthetic profiles can be calibrated and normalized to the continuum.

In order to calibrate the lines to the continuum of the FTS atlas, we multiplied the quiet Sun line intensity by a factor, such that the distance between the line and the FTS atlas is minimum at the continuum (see Figure~\ref{radyn_ibis_fts_qs}).

As mentioned in Section~\ref{sect:ibis} the H$\alpha$~and \ion{Ca}{2} 8542~\AA\ lines observed by IBIS had a wavelength width of 2 and 2.3~\AA\ with respect to the line center, not reaching the continuum. Therefore we fitted the available spectral range to the FTS atlas. The synthetic profiles were properly adjusted to the FTS continuum.

Our synthesized profiles for the quiet Sun are narrower and steeper than the observations, in agreement with \citet{2009ApJ...694L.128L}. As several authors \citep{2008A&A...480..515C, 2011A&A...528A.113D, 2012A&A...543A..34D} have previously explained, there are three principal reasons for this discrepancy:

\begin{figure*}[hbt]
\centering
  \epsscale{1}
  \plottwo{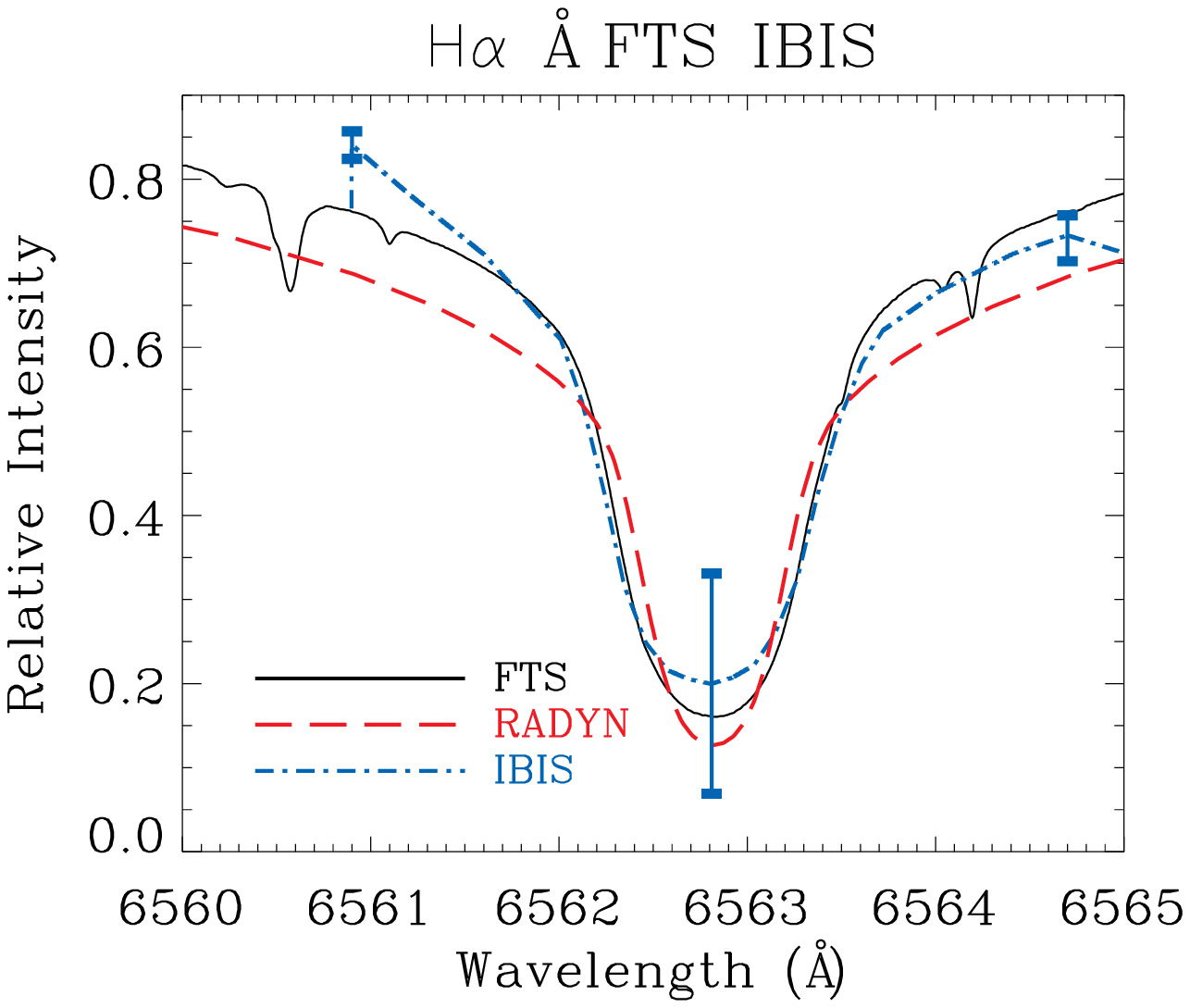}{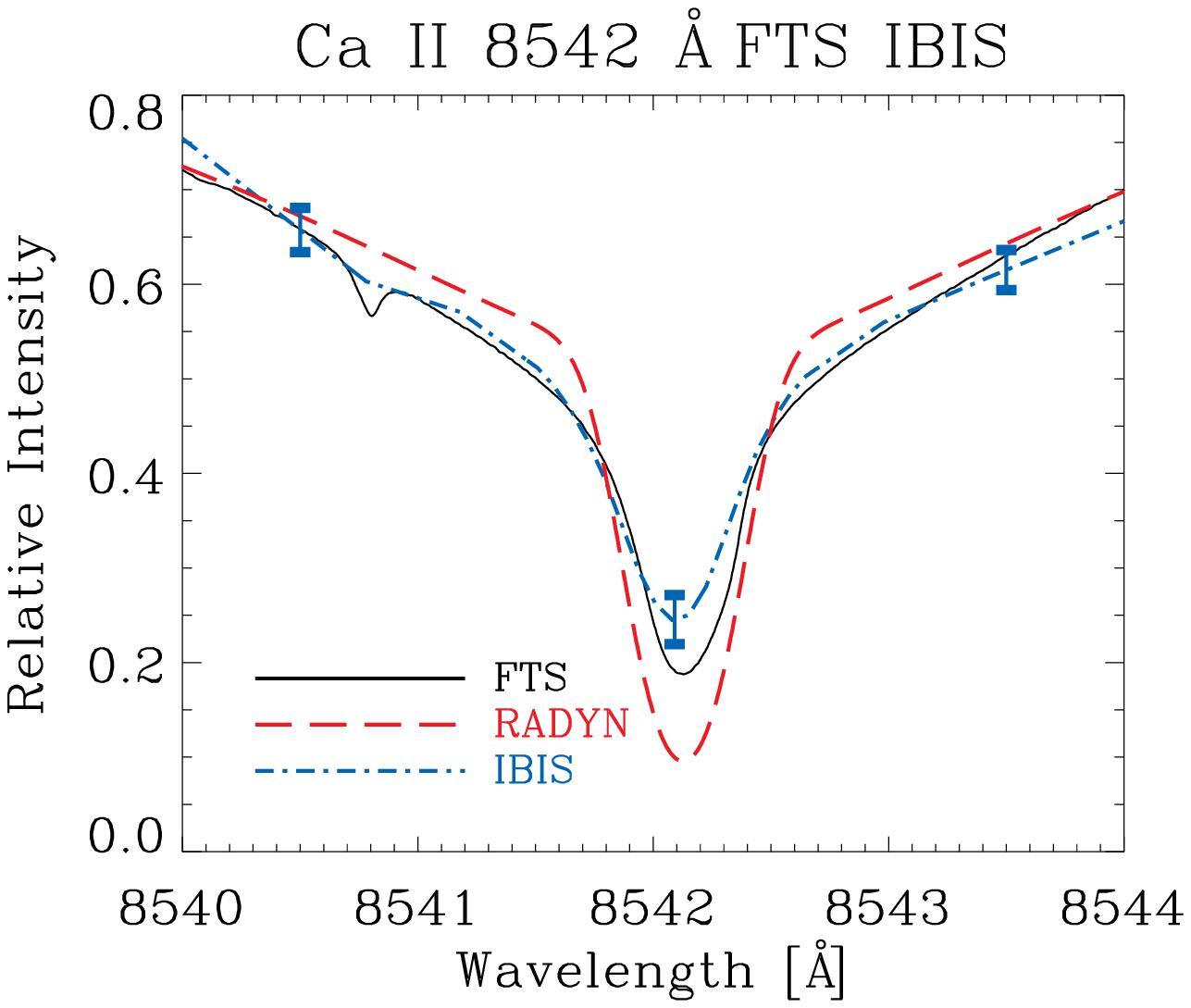}
  \caption{Comparison of the calibrated H$\alpha$~(left) and \ion{Ca}{2} 8542~\AA\ (right) synthetic (red dashed line) and IBIS ((blue dashed-dotted line) profiles for the quiet Sun taken at 19:22:40 and 19:22:15 UT respectively. The black solid line is the FTS atlas emission. The intensity scale is in relative units with respect to the FTS atlas continuum value. The error bars associated to the IBIS line profile represents the difference between both profiles during the flare at that particular wavelength position.}
  \label{radyn_ibis_fts_qs}
\end{figure*}

\begin{itemize}
\item In general, 1D simulation (even if they take into account the dynamics of the atmosphere) cannot catch all the structuring and small scales that are present in the chromosphere.

\item If the spatial resolution of the simulation is not high enough, the width of the average (spatio-temporal) profile is lower because the hydrodynamic simulations do not contain the necessary small-scale turbulence and the small scale motions are missing in the model. As \citet{2010MmSAI..81..576L} mentions, the increase in grid resolution causes an increase in amplitude of the velocity variations in the mid and upper chromosphere, and hence a widening of the average profile.

\item Because of higher opacities, synthetic lines usually show a much darker line core than the FTS atlas. This may be attributed to a low heating rate in simulations.
\end{itemize}

Figure~\ref{radyn_ibis_fts_qs} shows the resulting IBIS and RADYN lines fit to the FTS atlas after the calibration. The vertical error bars associated to the IBIS profile in Figure~\ref{radyn_ibis_fts_qs}(a) show the difference between the observed and the synthetic emission during the flare at different wavelength positions. The intensity difference between both H$\alpha$~lines in the wings is due to the poor fit of the IBIS H$\alpha$~profile to the continuum, because it is a very broad line and the observations covered only a wavelength range of 4~\AA.

After the quiet Sun profiles were calibrated and normalized to the continuum, we applied the normalization factor to the flaring line profiles and compared them in Figure~\ref{ha_ca_ibis_radyn} with the IBIS H$\alpha$~and \ion{Ca}{2} 8542~\AA\ observations H$\alpha$~and \ion{Ca}{2} 8542~\AA\ at 19:22:40 and 19:22:15 UT respectively.

For a better comparison of the shape of both line profiles, the synthetic profile has been shifted by 0.05~erg s$^{-1}$ cm$^{-2}$ \AA$^{-1}$ in order to align the wings of both lines. The line has been scaled to fit at the core, as shown by the green line of Figure~\ref{ha_ca_ibis_radyn}.

  \subsection{H$\alpha$~Line Profiles}\label{obs_radyn_ha}
\begin{figure*}[htb]
\centering
  \epsscale{1}
  \plottwo{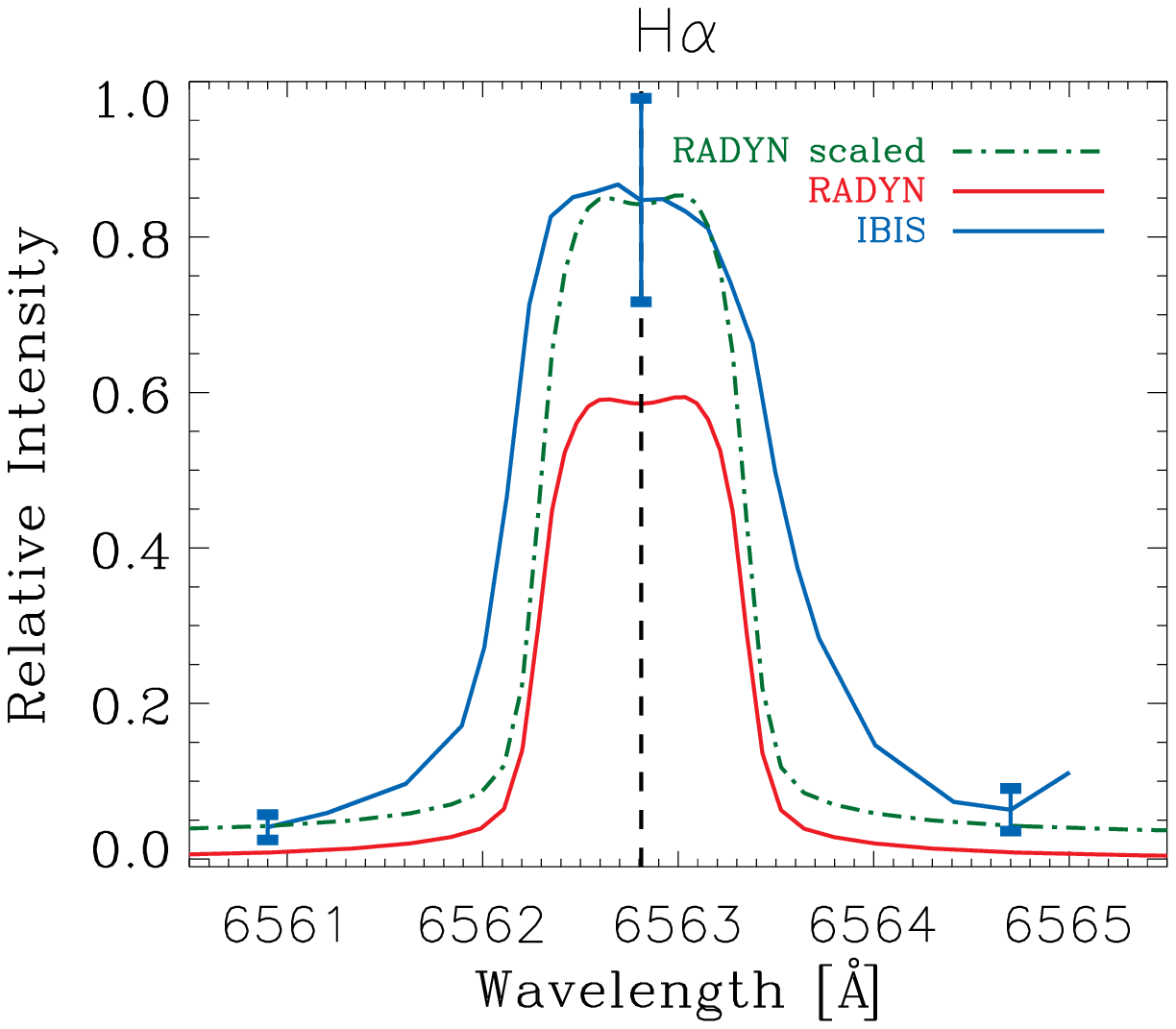}{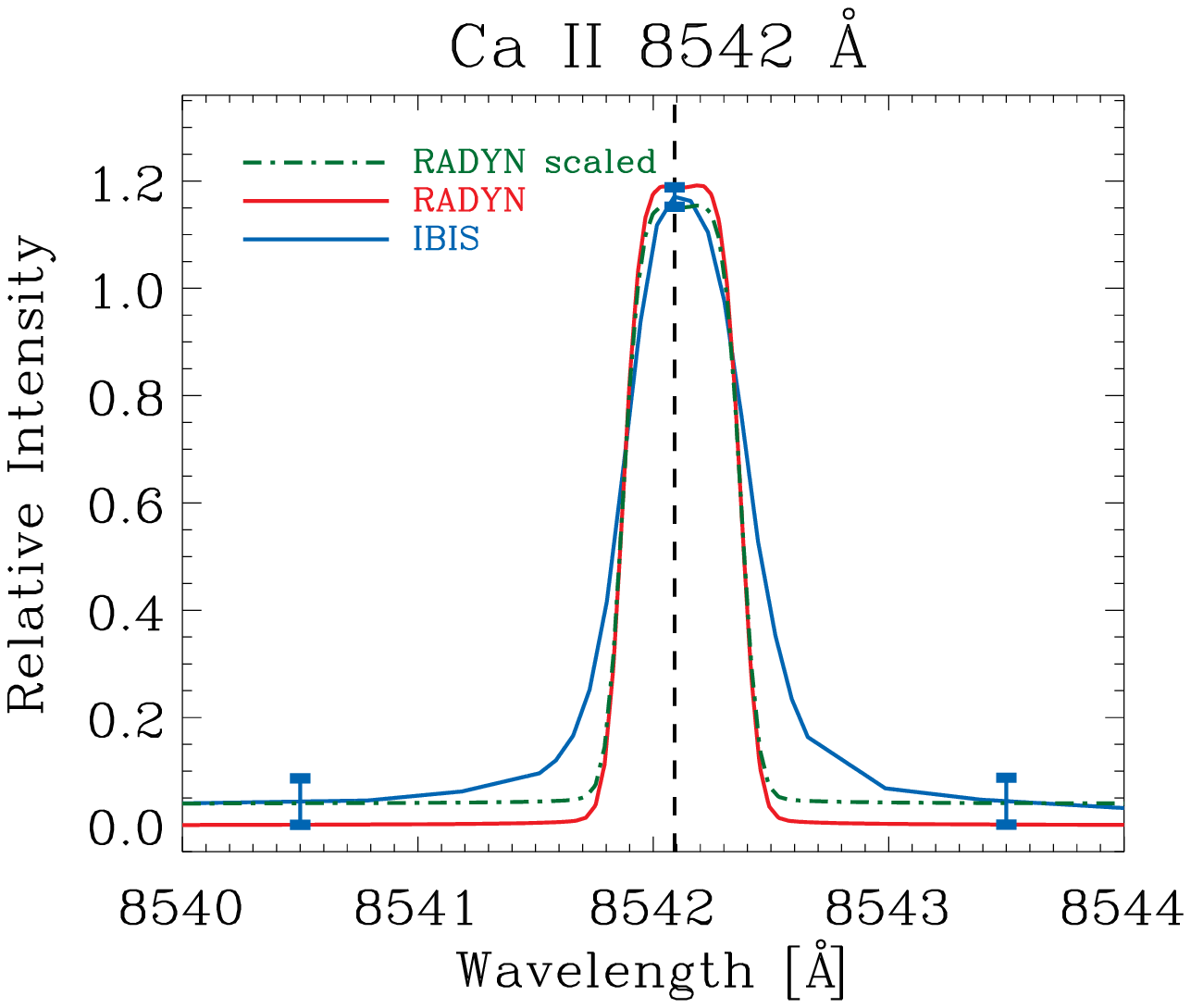}
  \caption{Comparison of the H$\alpha$~and \ion{Ca}{2} 8542~\AA\ lines observed by IBIS at 19:22:40 and 12:22:15 UT respectively (at the beginning of the gradual phase of the flare) with the synthetic ones, obtained from RADYN at the same time. The FWHM of the synthetic H$\alpha$~line is 1.07~\AA, while for \ion{Ca}{2} 8542~\AA, it is 0.51~\AA. The green line is the synthetic profile scaled at the core and the continuum of the line. The intensity scale is in relative units with respect to the FTS atlas continuum value. The vertical dashed line indicates the line center. The error bars associated with the IBIS line profile represent the difference between observed and synthetic profiles during the flare at that particular wavelength position.}
  \label{ha_ca_ibis_radyn}
\end{figure*}

Figure~\ref{ha_ca_ibis_radyn}(a) compares the H$\alpha$~excess line profile obtained from RADYN (red) and observed by IBIS (blue) from 19:22:40 to 19:23:07 UT. As mentioned in Section~\ref{sect:calibrate_qs}, the intensity shift between both profiles at the wings of the line is due to the uncertainty of the continuum fit. 

The core of the simulated line is $\approx$ 23\% less bright than the observation and the wings are slightly narrow, as discussed in Section~\ref{sect:calibrate_qs}. The core of H$\alpha$~is sensitive to the temperature pattern due to the low mass of the hydrogen atom \citep{2012ApJ...749..136L} which can contribute to the difference in the core emission. The green profile in Figure~\ref{ha_ca_ibis_radyn}(a) is the synthetic profile scaled to fit at the core and at the continuum of the IBIS profile and the vertical error bar represents the difference between the observed and the synthetic emission at different wavelength positions, calculated for this time.

	The standard NLTE line formation assumption of statistical equilibrium does not properly fit the observations because of the slow collisional and radiative transition rates when compared to the hydrodynamical timescale. In addition, at least the Lyman-$\alpha$~and $\beta$~lines need to be modeled with partial frequency redistribution (PRD) because of their strong influence on the H$\alpha$~line \citep{2010MmSAI..81..576L}. Inclusion of these effects might significantly increase the H$\alpha$~opacity. Thus, proper modeling of the H$\alpha$~line requires full time-dependent radiative transfer with PRD in tandem with the hydrodynamic evolution, a Herculean task that has not been done in 1D hydrodynamic simulations since so far all 1D simulations perform radiative transfer assuming complete redistribution \citep{2005ApJ...630..573A, 2009A&A...499..923K, 2010ITPS...38.2249V}.

As \citet{2012ApJ...749..136L} explains, PRD effects can be approximated by truncating the Lyman line profiles. RADYN truncates the Lyman line profiles at $\pm$~64~km~s$^{-1}$ away from the line center frequency. This is a reasonable approximation, with the effect that we get higher heights of formation than if we assumed PRD.

   \subsection{\ion{Ca}{2} 8542~\AA\ Line Profiles}\label{obs_radyn_ca}
Figure~\ref{ha_ca_ibis_radyn}(b) compares IBIS (blue line) and RADYN (red line) at the time range 19:22:15 - 19:22:42 UT, showing a good agreement between both profiles. The green line profile is the synthetic profile scaled to fit at the core and at the continuum of the IBIS profile.

As mentioned by \citet{1987A&A...181..103S}, assuming complete spectral redistribution for the scattered photons may be a poor approximation for the \ion{Ca}{2} 8542~\AA\ and this could explain the increased intensity in the core. \citet{2009ASPC..415...87L} and \citet{2009ApJ...694L.128L} investigated the formation of \ion{Ca}{2} 8542~\AA\ in 3D MHD models that extended up into the corona, finding that the 3D effects are important, especially in the core of the line. Considering these two statements and that both lines differ in the core in $\approx$ only 2.4\%, our synthetic \ion{Ca}{2} 8542~\AA\ profile fits very well to the observations. As discussed in Section~\ref{sect:calibrate_qs}, the wings area narrower in the simulated profile because of a lack of small scale dynamics in the model.

From the study of the monochromatically optical depth of ~\ref{eq_contribution_function}, $\tau_{\nu}=1$, (green line in Figure~\ref{evol_ci_ha_ca}) we get that the \ion{Ca}{2} 8542~\AA\ is formed between 0.15~Mm in the wings and 1.05~Mm in the core, while H$\alpha$~is formed at 0.2~Mm in the wings of the line and 1.15~Mm at the core. Even if both lines had almost a similar formation height range and the same atmospheric conditions, \ion{Ca}{2} 8542~\AA\ fits better to the observations than H$\alpha$. Since the 3d $^2$D$_{3/2,5/2}$ levels are metastable, they can only be populated from below by collisional excitation, strengthening the sensitivity of the \ion{Ca}{2} infrared triplet to local temperature. On the other hand, the lower energy level of H$\alpha$~is 10 eV higher than \ion{Ca}{2} 8542~\AA, where the population is very sensitive to the atmospheric parameters, explaining the different behavior of the two lines.

   \section{Summary and Discussion}\label{sect:conclusions}
We have presented in this paper a self-consistent, data-driven radiative hydrodynamic simulation of an M3.0 flare and its comparison with high-resolution spectroscopic observations by the IBIS instrument. By fitting X-ray spectra of this flare observed by {\it RHESSI}, we inferred the flux of the non-thermal accelerated electrons, which was used as an input to the RADYN code. The RADYN code incorporates careful treatments of atomic and molecular physics together with radiative transfer and hydrodynamics, allowing us to study the evolution of the flaring atmosphere as well as the detailed chromospheric emission. We synthesized the H$\alpha$~and \ion{Ca}{2} 8542~\AA\ excess line profiles, which in general are in agreement with those observed by IBIS. Specifically, as shown in Figure~\ref{ha_ca_ibis_radyn}, the synthesized \ion{Ca}{2} 8542~\AA\ emission is consistent with the observations within the expected uncertainties, while the H$\alpha$~synthetic line is $\approx$ 23\% fainter in the core than the observations. Both synthetic lines have similar shapes as the observed line, but the synthetic lines exhibit a typical flattening in the core due to an overestimate of the opacity, as discussed in Section~\ref{sect:calibrate_qs}.

There are several limitations in our approach, which could be improved in the future. For example, solar flares are complex and dynamically three-dimensional in nature. The current 1D models are not yet capable of handling this properly and solving the equations of non-equilibrium and non-LTE optically thick radiative transfer in multiple dimensions. The inclusion of a quasi thermal component in the electron distribution, in addition to the non-thermal component, and a proper treatment of the electron transport process can play an important role in the estimation of the electron heating rate \citep[e.g.,][]{2009ApJ...693..847L}. Moreover, taking into account the uncertainties in the {\it RHESSI} fitting parameters and the measurement of the area may reduce the differences between the synthesize and observed line profiles.

As noted by \citet{2010MmSAI..81..576L}, the dominant chromospheric energy loss on the quiet Sun is through radiation in strong lines, and a comprehensive model of the chromosphere cannot be constructed without the inclusion of the underlying photosphere and upper convection zone and the overlying lower corona. Our RADYN code lacks detailed treatment of the photospheric radiation, which could potentially contribute to the small discrepancies between the synthetic and observed lines. A better reproduction of the observations may require higher resolution, larger computational domains, an improved treatment of radiation and non-equilibrium hydrogen ionization. To improve the modeling of the hydrogen transitions and in particular H$\alpha$, 3D NLTE time-dependent radiative transfer codes including PRD would be a considerable undertaking.

By increasing the flux of the injected electrons a factor of two, the atmosphere evolves faster at initial times and as result the line profiles show red-wing asymmetries at early times. During the impulsive phase of the flare, the H$\alpha$~emission increases by a factor of 1.2 and \ion{Ca}{2} 8542~\AA\ by factor of 1.5. At later times the lines are fainter, decreasing by a factor of 1.3 at 19:23:52 UT for H$\alpha$~and a factor of 1.1 for \ion{Ca}{2} 8542~\AA, mostly because the flattening at the line core becomes more obvious. There is still a mismatch at the wings of the lines, which are still narrower than the observations.

Extending the analysis to a multiwavelength study as the one presented by \citet{2014ApJ...793...70M} demonstrates the value of bringing together observations over a broad spectral range. The combination of ground-based observations, such as H$\alpha$~and \ion{Ca}{2} 8542~\AA\ with the recently launched {\it Interface Region Imaging Spectrograph} \citep[{\it IRIS};][]{2014SoPh..289.2733D} would give a more detailed information about the response of the chromosphere during a solar flare.

\acknowledgments
The authors thank M. Carlsson for the stimulating discussions. Work performed by F.R.dC., V.P. and W.L. is supported by NASA grants NNX13AF79G and NNX14AG03G. 
L.K. and A.S.D. are supported by NASA LWS grant NNX13AI63G.

\bibliographystyle{apj}
\bibliography{ads}
\end{document}